\documentclass[%
 preprint,
 amsmath,amssymb,
 aps, physrev,
]{revtex4-2}

\usepackage{graphicx}
\usepackage{dcolumn}
\usepackage{bm}

\begin{document}

\preprint{}

\title{\textbf{A dynamical memory with only one spiking neuron} 
}%

\author{Damien Depannemaecker*}
 \affiliation{Aix Marseille Universit\'e, INSERM, INS, Institut de Neurosciences des Syst\`emes, Marseille, France}
\author{Adrien d'Hollande*}
\affiliation{Universit\'e Paris-Saclay, CNRS
Laboratoire de Physique des Solides, 91405, Orsay, France}
\author{Jiaming Wu}
\affiliation{Universit\'e Paris-Saclay, CNRS
Laboratoire de Physique des Solides, 91405, Orsay, France}
\author{Marcelo J. Rozenberg}
\affiliation{Universit\'e Paris-Saclay, CNRS
Laboratoire de Physique des Solides, 91405, Orsay, France}
\affiliation{Universit\'e Paris-Cité, CNRS
Integrative Neuroscience and Cognition Center, 75006, Paris, France}






\date{May 20, 2025}

\begin{abstract}
Common wisdom indicates that to implement a Dynamical Memory with spiking neurons 
two ingredients are necessary: recurrence and a neuron population.
Here we shall show that the second requirement is not needed. We shall demonstrate
that under very general assumptions a single recursive spiking neuron can realize
a robust model of a dynamical memory.
We demonstrate the implementation of a dynamical memory in both, software and hardware.
In the former case, we introduce trivial extensions of the popular aQIF and AdEx models. 
In the latter, we show traces obtained in a circuit model with a recently proposed 
memristive spiking neuron.
We show that the bistability of the theoretical models can be understood in terms of a self-consistent problem that can be represented geometrically.
Our minimal dynamical memory model provides a simplest implementation of an important neuro-computational primitive, which can be useful in navigation system models based on purely spiking dynamics.
A one neuron dynamical memory may also provides a natural explanation to the surprising recent observation that the excitation bump in Drosophila's ellipsoidal body is made
by just a handful of neurons.
\end{abstract}

\maketitle
*co-first authors

\section{\label{sec:intro}Introduction:\\ The challenge of a stable dynamical memory\protect}
Minimal models have the beauty of simplicity and the relevance of conceptual clarity. 
In neuroscience, dynamical memory, also called working memory, is a key unit of minimal cognitive behavior. 
Its function is to keep a spiking state across time, i.e.
it implements neural persistent behavior. It is intuitive that such a persistence can be achieved by a reverberating state.
However, conventional wisdom also indicates that a single spiking neuron cannot realize a stable self-excitatory
state. If a neuron feeds back its spikes with a self-excitation, this would increase the excitation and further
increase the spike rate. Hence, following this intuition the basic models of persistent activity or ``working memory''
are though to require of neural networks where the mutual excitation between neurons is counterbalanced by suitable 
inhibitory synaptic couplings \cite{gerstner,wilson}. Thus, dynamical memory models are based on neural populations with
fine tuning between these two opposite 
interactions and presenting hysteresis 
effects in their response \cite{seung-autapse,seung-stability,brunel,koulakov2002model}.

In contrast to this conventional wisdom, we shall demonstrate here that a minimal dynamical memory  
requires neither a population nor fine-tuning,
and that a single self-excited neuron is sufficient. We shall show that the spiking state of the 
minimal motif can be posed as the solution of a self-consistent problem. This will allow us to see that
the intuition of an unstable runaway state is true, but it will also reveal the existence of a second 
solution that is stable and can be realized under quite general assumptions. 

Rather surprisingly, this basic dynamical behavior seem to have been missed 
in theoretical studies so far. 
We shall also demonstrate the actual implementation of the minimal dynamical memory in analog hardware,
adopting a memristor based realization of a spiking neuron \cite{wu2023bursting,bookchapter} 
with a self-excitatory membrane current \cite{wu2024neurosynaptic}. This further demonstrates the
generality of our model. Moreover, it opens the way for potential applications, as it provides
a physical embodiment of a relevant computational primitive for robotics and 
spiking neural networks for artificial intelligence.

\section{\label{sec:results}Results\protect}
\subsection{\label{sec:self-spiking}A self-consistent spiking-frequency problem\protect}
Let's show how finding the spiking state of single self-excited neuron can be cast as a self-consistent problem. 
We begin by presenting general model equations for a spiking neuron and a membrane current that is modulated by the neuron's spiking state. Perhaps the two most popular theoretical models are the one introduced by Izhikevich \cite{Izhikevich2003-ed} and the AdEx \cite{Brette2005-bb}.
They can be written in a general form,
\begin{eqnarray}
    \label{eq:neuron_model_1}
    C \frac{dV}{dt} &=& -g_L V + F(V) - I_w + I_{ext} \\
    \tau_w \frac{d I_w}{dt} &=& - I_w +  H(V)
    \label{eq:neuron_model_2}
\end{eqnarray}

where $V$ denotes the membrane potential of the neurons, $C$ is the capacitance and $g_L$ the leak
conductance. $I_0$ is the input excitatory current and $I_w$ denotes the intrinsic membrane current. 
$F(V)$ is a non-linear function of $V$ which captures the spike initiation phenomenon (i.e. non-linear subthreshold behavior close to the threshold, biophysically associated with the opening of sodium voltage-gated channels)  
and $H(V)$ specifies the coupling of the $I_w$ slow variable equation with that of the
membrane potential. For aQIF, Izhikevich and AdEx, $H(V)$ adopts a simple linear form, but it can also be 
non-linear, such as in the Conductance-Based Adaptive Exponential Integrate-and-Fire model (CAdEx) \cite{Gorski2021-aa}.

The main difference between those models is in the definition of the non-linear function $F(V)$, but 
nevertheless they exhibit
a similar qualitative behavior. In the aQIF and Izhikevich model $F(V)$ is a quadratic polynomial, while
in AdEx is an exponential function. In these models, the initiation of the spike occurs when the
function $F(V)$ becomes positive enough to overcome the (negative) leak term and causes a runaway
of the membrane potential $V(t)$. Hence an {\it ad hoc} cut-off is introduced which is known as the
reset condition and reads,
\begin{eqnarray}
    {\rm if}\ \ \ \  V = V_{spike} \ \ {\rm then}\ \ \  V \leftarrow V_{reset}  
    \label{eq:cut-off}
\end{eqnarray}
where $V_{spike}$ and $V_{reset}$ are parameters of the model.

The emission of the spike also has an effect on the intrinsic current $I_w$. Then, when $V$ reaches $V_{spike}$ the current $I_w$ is increased in an amount given by the model parameter $b$. Hence,
\begin{eqnarray}
    {\rm if}\ \ \ \  V = V_{spike} \ \ {\rm then}\ \ \  I_w \leftarrow I_w + b  
\label{eq:current}
\end{eqnarray}

\begin{figure}
    \centering
    \includegraphics[width=0.9\linewidth]{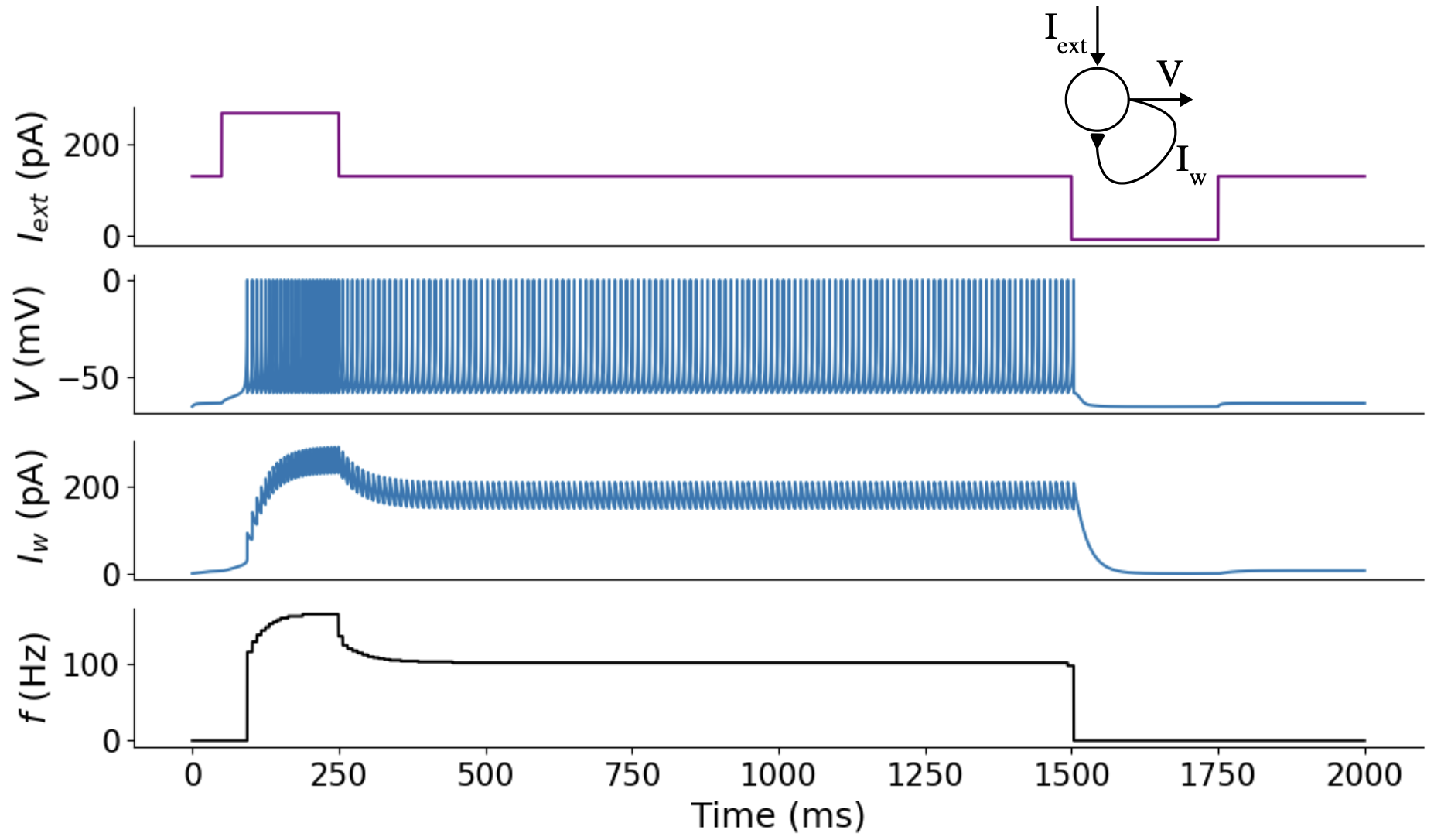}
    \caption{ On top right, motif of a spiking neuron with an intrinsic
    self-excitatory current. Top to bottom panel show the bi-stable dynamical behavior of a working memory. 
    Until t = 100s a subthreshold input current $I_{ext}=I_0$ = 130pA is applied and the
    neuron is quiescent.
    An above-threshold current pulse of 140 pA starts the spiking ($I_{ext}=I_0+140$pA) . 
    The self-excitatory feedback $w$ builds-up and produces a fast increase of the
    frequency $f$ , which does not runaway but saturates. When the initial pulse is terminated 
    at t = 250s the input current returns to the subthreshold value, but the 
    action potential $V$ spiking continues. At t = 1500s the input current is set to 
    zero ($I_{ext}=0pA$) during 250ms (inhibitory pulse) and the spiking activity decays back to the 
    quiescent state. The model parameters
    are $C$=200pF, $g_L$ = 10 nS/mV, $E_L$ = -65mV, $V_T$ = -55mV, $\Delta_T$ = 2mV,
    $\tau_w$ = 20ms, $a$ = 4nS, $b$ = 60pA, $V_{reset}$ = -58mV.}
    \label{fig:1}
\end{figure}

This last equation completes the definition of the models, which is given by the system of Eqs.~(\ref{eq:neuron_model_1}) to (\ref{eq:current}).
In Fig.~\ref{fig:1}. we show a schematic representation of the system, which emphasizes that the intrinsic membrane
current can be viewed as a self-recurrent coupling.

In the introduction above, we mentioned that a dynamical memory is assumed to require 
excitatory recurrence to gain persistence. 
This motivates us to extend the above neuron models by turning the 
adaptation variable into a self-excitatory one. 
This is simply done by changing {\it a single sign} in
the $I_w$ term of Eq.~(\ref{eq:neuron_model_1}), which now reads,
\begin{eqnarray}
   C \frac{dV}{dt} &=& -g_L V + F(V) + I_w + I_{ext} \\
    \tau_w \frac{d I_w}{dt} &=& - I_w +  H(V)
   \label{eq:neuron_model_with_ex}
\end{eqnarray}
We should mention here that excitatory membrane currents do exist in biological neurons, such as the
transient inward Ca2+ ionic current, known as $I_T$, or the Ca-activated K currents $I_{KCa}$.

In the following, we shall explore the behavior of the excitatory version of these popular 
theoretical models, defined by the Eq.~(\ref{eq:neuron_model_with_ex}), supplemented by Eqs.~(\ref{eq:cut-off}) and  (\ref{eq:current}). For definiteness, in what follows, we shall adopt for $F(V)$
that of the aQIF model \cite{Izhikevich2003-ed,gerstner}  (also equivalent under certain conditions to 
Izhikevich). 
The specific interest to adopt the aQIF is that it has been shown that an exact mean-field model of 
these neurons can be obtained \cite{montbrio,chen-campbell}. 
This is known as a neural mass, which is a relevant feature of the so called whole brain
network models \cite{virtual_brain}, where the mean-fields represent brain nodes.
For the sake of completeness, and to demonstrate the generality of the minimal dynamical memory formulation
we consider the extensions of Izhikevich and AdEx models in the Appendix, and show that it exhibits similar properties. 

Thus, we shall study next the extension of the aQIF 
model, which we term SEQIF (for Self-Excitatory Quadratic Integrate and Fire).
The model equations are defined below,
\begin{eqnarray}
 \label{eq:eqif}
   C \frac{dV}{dt} &=& g_L (E_L - V) (V_T - V) + I_w + I_0 \\
  \label{eq:leaky-integrate}
    \tau_w \frac{d I_w}{dt} &=& a(V-E_L) - I_w   \\
    \nonumber\\
    \label{eq:model}
      {\rm if}\ \ \   V &=& V_{spike} \ \ {\rm then}\ \nonumber \\ 
      V &\leftarrow & V_{reset} \ \ \ {\rm and} \ \ \  I_w \leftarrow I_w + b   
\end{eqnarray}
where $a$, $E_L$ and $V_T$ are parameters of the model.

\begin{figure}[h]
    \centering
    \includegraphics[width=1\linewidth]{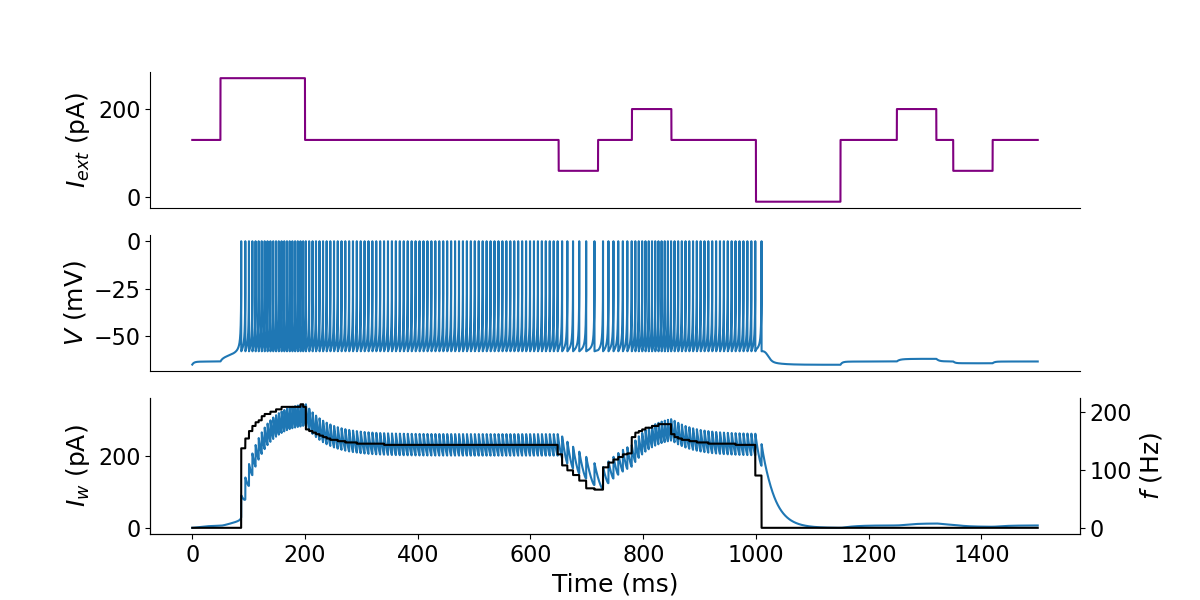}
    \caption{Stability of the dynamical memory attractor. The traces show the behavior
    of the dynamical memory under perturbations. A first pulse at 100ms drives the
    neuron to the dynamical attractor. After the pulse the spiking persists although
    the $I_{ext}$ is subthreshold. After $t$=700ms a sequence of perturbation pulses 
    is applied demonstrating the remarkable stability of both, the dynamical attractor
    and of the quiescent state. The bottom panel demonstrates the approximate linear relation
    between the spiking frequency and the feedback current (cf.\ Eq.~(\ref{eq:lin})).
    }   
    \label{fig:2}
\end{figure}

In the left panel of Figs.~\ref{fig:1}. and \ref{fig:2}. we present the key 
finding of the present work, namely that
our extension of the popular self-adaptation models to the case of self-excitation 
can implement a minimal working memory

We shall now show how the dynamical self-sustained spiking state can be
understood as a self-consistent solution by means of a geometrical construction
To this end we need two functions that characterize the system: the neuron's response function and
the intensity of the synaptic or membrane current that it generates when it is spiking at a given
frequency $f$. 

\begin{figure}
    \centering
    \includegraphics[width=0.8\linewidth]{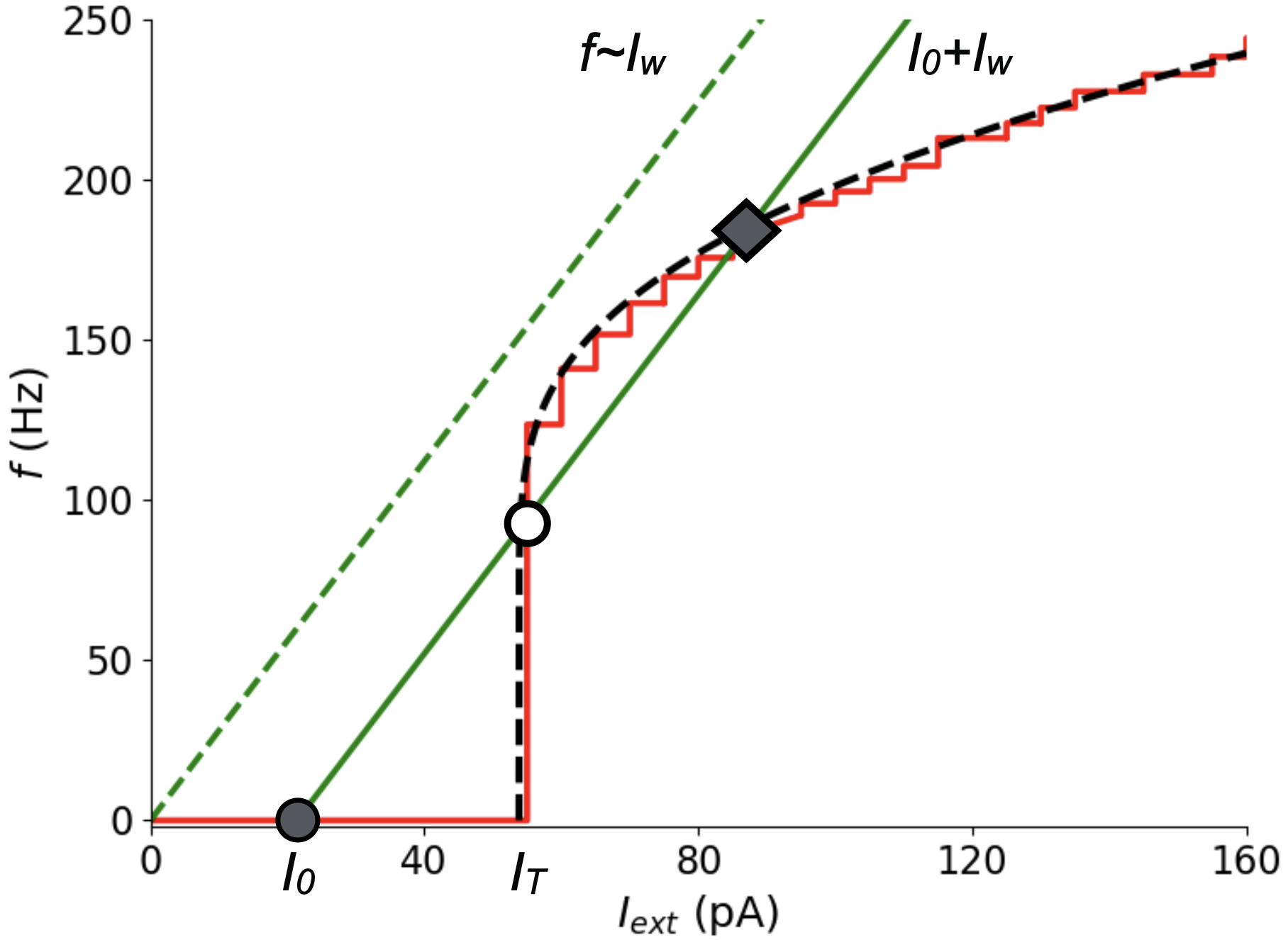}
    \caption{Response or activation function $f(I_{ext})$ computed without feedback
    (red). Activation threshold current $I_T$. Subthreshold current $I_0$. The critical LIF 
    activation form $\sim 1/\{1-log[(I-I_T)/I_T]\}$ provides a good fit \cite{izhikevich-dns-book} 
    (black dashed).
    The feedback current $I_w(f)$ is approximately linear (cf.\ Fig.~\ref{fig:2}. (green
    dashed). The total input current, subthreshold plus feedback $I_0+I_w$ (green).
    Two self-consistent attactors: a stable fixed-point (black circle) and the stable
    dynamical attractor (black square) separted by an unstable fixed-point (white circle).
    }   
    \label{fig:3}
\end{figure}

The response function is obtained numerically by exciting the neuron with a constant input 
current $I_{ext}$
and in the absence of adaptation or self-excitation feedback, i.e. setting $I_w = 0$ in Eq.~(\ref{eq:eqif}).
We obtain the response function $f(I_{ext})$ that is shown in Fig.~\ref{fig:3}. for the SEQIF model,
which indicates that there is a threshold current $I_T$ required to elicit spikes.
In the figure we can also observe that this activation function is well captured by
the LIF model expression \cite{izhikevich-dns-book}.

On the other hand, we also need the frequency-dependent membrane current $I_w$, which 
can be easily obtained analytically.
From Eqs.~(\ref{eq:leaky-integrate}) and (\ref{eq:model}) we see that each spike contributes to the 
current $I_w$ by an amount $b$,
and its effect has a duration given by $\tau_w$.
Hence, if the spikes are emitted with a frequency $f$ and the 
interspike-interval (ISI $ = 1/f$) is 
smaller than $\tau_w$, the current can be approximated (up to a geometrical factor of order 1) by,
\begin{equation}
    \label{eq:lin}
    I_w = (b\tau_w)f
\end{equation}
Thus the membrane current and the frequency are linearly
related, as shown in Fig.~\ref{fig:3}., where for convenience we plotted the inverse relation
$f(I_w)$.

\begin{figure}[t]
    \centering
    \includegraphics[width=1\linewidth]{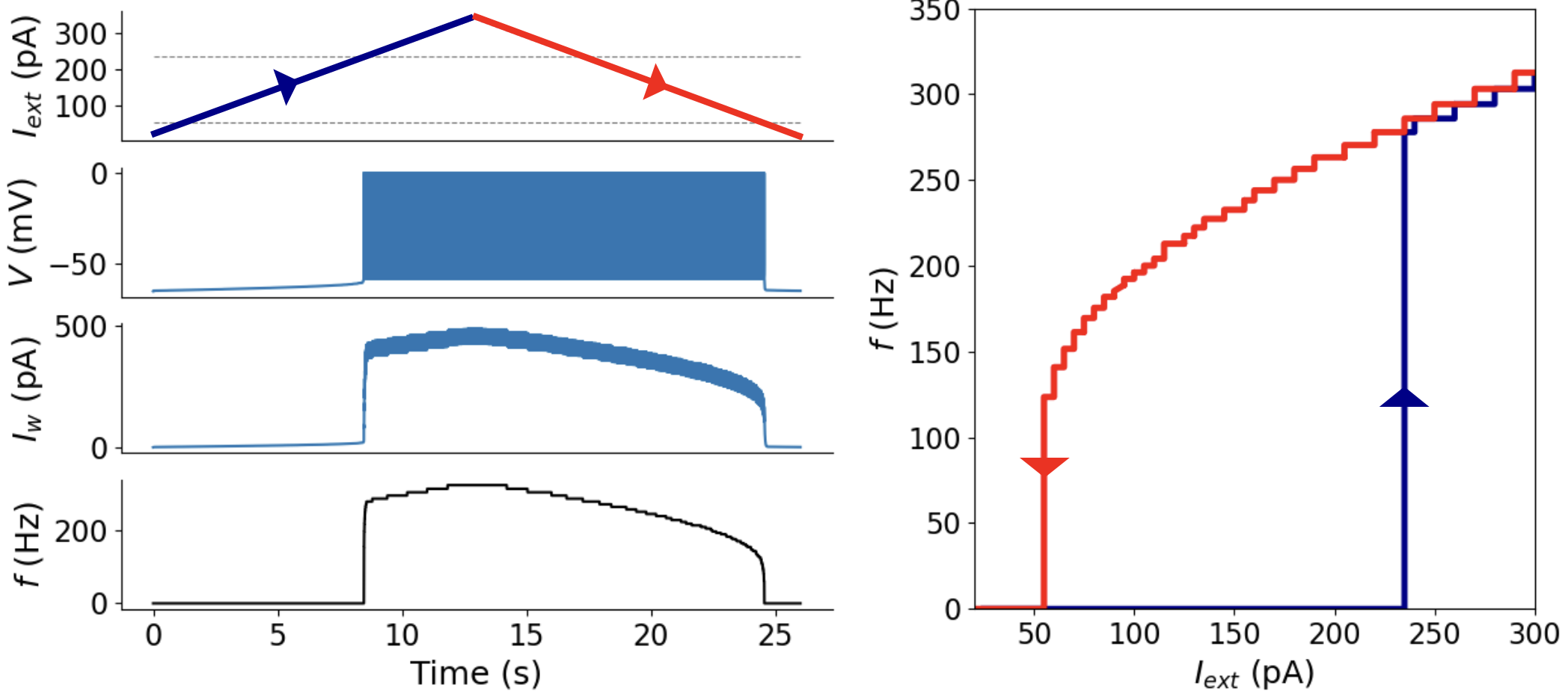}
    \caption{Left: behavior of the SEQIF model when it is excited by a gradually increasing (blue)
    and then decreasing (red) input current $I_{ext}(t)$ shown on the top panel. Below, the spiking response action
    potential $V$, and then
    the feedback membrane current $I_w$. The bottom panel shows the spiking frequency.
    Right: Hysteretic behavior of the response function $f(I_{ext})$ obtained increasing (blue) and decreasing (red)
    external input current. 
    }   
    \label{fig:4}
\end{figure}

We can now pose the problem of obtaining the spiking state of the neuron as follows:
We need to find the spike frequency $f^*$ that will produce a membrane current
$I_w(f^*)$, such that $f^* = f(I_0+I_w(f^*))$. This is a self-consistent condition that 
can be represented geometrically as shown in Fig.~\ref{fig:3}. The solution to the 
self-consistent state is given by the intersection of $f(I_{ext})$
and $f(I_0+I_w)$. As can be seen in the figure, there are two crossing points, shown by
circles. However, the white one is unstable and the black one is stable. This can be
easily seen by considering the evolution of a small perturbation.

Interestingly, the existence of the stable solution seems to have been missed so far,
as we were not able to find it in textbooks nor the literature. The reason may be
that intuition indicates that the self-excitation of spikes should be a run-away
behavior, which indeed corresponds to the first unstable fixed point. In fact,
in general any neuron model it is present and in the simple ReLU is the only
one. In contrast, the stable fixed point would be found in any neuron model
where the activation function is convex or where the membrane current show
a saturation effect with the frequency, i.e. $I_w(f)$ is sub-linear.

We now turn to the response function of the full neuron
model, including the excitatory feedback (cf.\ left panel of Fig.~\ref{fig:1}.). 
As one may expect of a memory functionality,
the response function exhibits hysteresis. To see this, we may gradually increase the 
intensity of the applied current $I_{ext}$ until the neuron starts spiking and
the decrease until the quiescent state is found again.
The results are shown in Fig.~\ref{fig:4}.

The activation shows a strong hysteresis effect, consistent with the remarkable stability
as was previously discussed. It is interesting to note that in the increasing current
branch (blue) the self-feedback
modifies the both, the value of the threshold and the activation behavior 
from type I to type II. In contrast, along the decreasing current branch (red) the
neuron model remains of type I and the threshold also remains unchanged.
This is easy to see, since $I_w \to 0$ as $f \to 0$.

\subsection{\label{sec:nullb}Nullcline and bifurcation\protect}

\begin{figure}[t]
    \centering
    \includegraphics[width=1\linewidth]{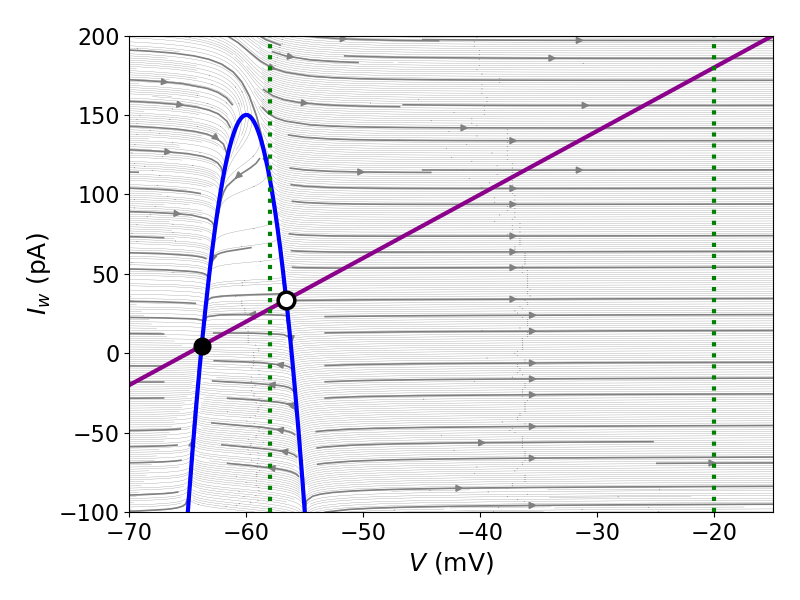}
    \caption{
    Phase-plane of the SEQIF model for $I_{ext} = 0$. In blue the $V$-nullcline and in magenta $I_w$-nullcline, crossing in 
    two fixed-points, a stable (black) associated with resting state and an unstable (white). On the right side of the 
    right branch of the $V$-nullcline the flow diverges in the direction to the spiking threshold ($\approx$-20mV) shown by a green dotted line. 
     The reset potential ($\approx$-58mV) is also shown by a green dotted line.
     }
    \label{fig:5}
\end{figure}

The  dynamics described previously can be understood by considering the topology of the phase plane shown in Fig.~\ref{fig:5}. 
The $I_w$-nullcline is linear and the $V$-nullcline take the quadratic form of the non-linearity of this model, enabling the existence of two fixed points: one stable corresponding to the resting state, and one unstable on the separatrix following the right branch of the $V$-nullcline. The flow diverges, directing the system towards the reset threshold.   

\begin{figure}[t]
    \centering
    \includegraphics[width=1\linewidth]{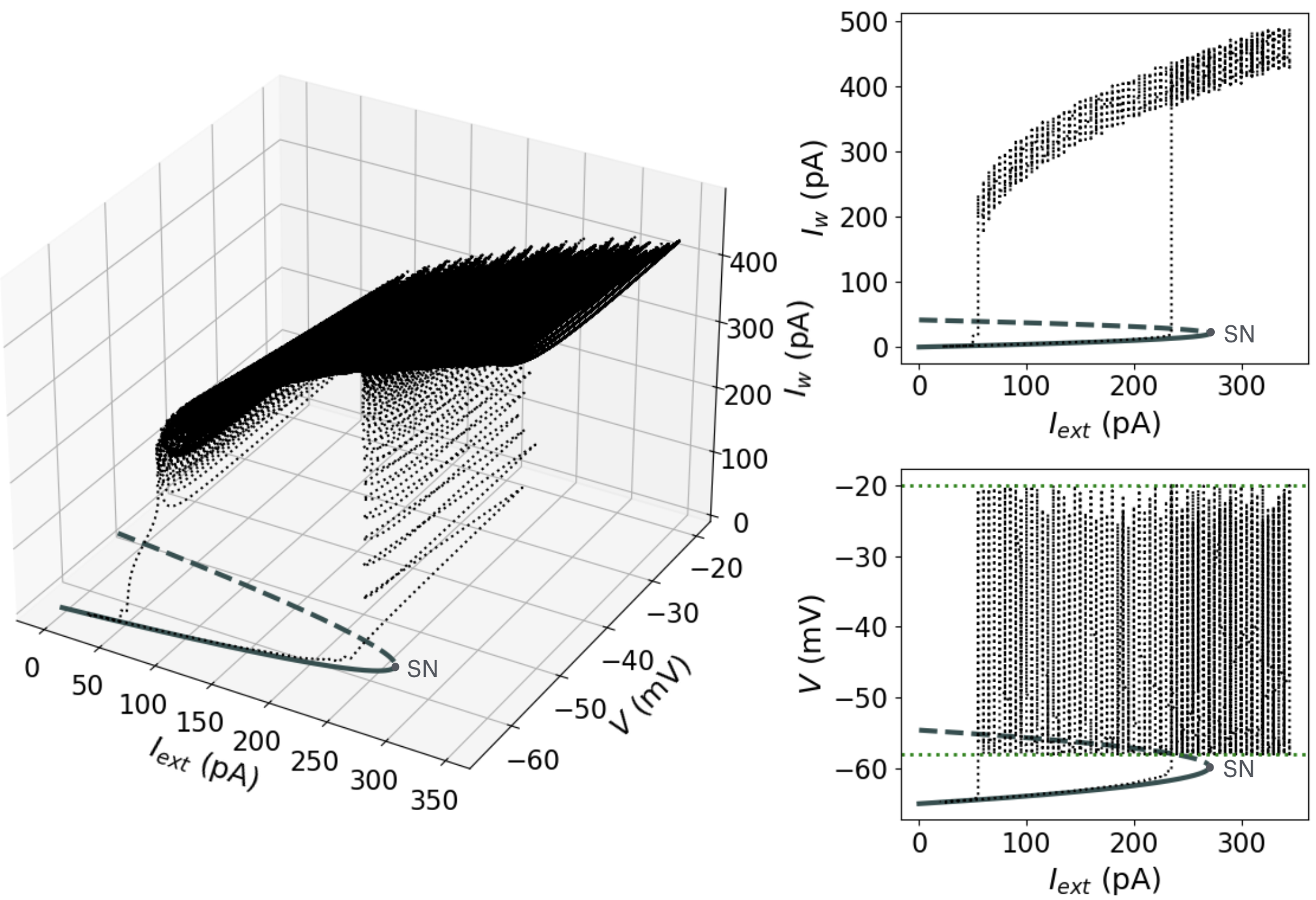}
    \caption{On the left: Stable fixed-point (dark gray line) and unstable fixed-point (dashed dark gray line) disappearing in a saddle-node bifurcation 
    while increasing $I_{ext}$. Within a range of $I_{ext}$ the stable fixed point co-exist with a limit-cycle defined by the reset mechanisms, as shown 
    by the dotted line, obtain from the simulation shown in Fig.~\ref{fig:4}. On the right: On top projection on the $I_{ext} - I_w$ plane, here the width 
    of the limit-cycle correspond to the parameter $b$. Below projection on the $I_{ext} - V$ plane, where the width of the limit-cycle is constraint 
    by $V_{spike}$ and $V_{reset}$ (shown by the two green dashed lines).}    
    \label{fig:6}
\end{figure}

The position of the limit cycle and the coordinates of the two fixed-points can be studied depending on $I_{ext}$, as shown in Fig.~\ref{fig:6}. 
As $I_{ext}$ increases, the system transitions from a regime with a single stable fixed point to a bistable regime where both a stable fixed point 
and a stable limit-cycle coexist. This bistability arises due to the nonlinear dynamics introduced by the reset condition, which effectively extends 
the two-dimensional phase space into a hybrid dynamical system. Indeed, This reset mechanism is equivalent to a third dimension enabling the existence 
of a limit-cycle, corresponding to the spiking activity. The limit-cycle does not emerge from the continuous flow of the differential equations alone, 
but is instead a consequence of the imposed discontinuity in the voltage dynamics. Along the $V$ direction, its location is fixed and constrained 
by the model parameters $V_{spike}$ and $V_{reset}$ (shown with green dotted lines in Figs.\ref{fig:5} and \ref{fig:6}). 
Along the $I_{w}$ dimension, the width of the limit cycle depends on the model parameter $b$, and its steady state depends on the dynamics of $I_{w}$, particularly on the time constant $\tau_w$.. 

The presence of this artificial, yet physiologically motivated, reset condition ensures that the system cannot settle at the unstable fixed point once the 
threshold is crossed. Instead, it follows a repeatable path that mimics sustained spiking activity. This hybrid structure, a continuous flow interrupted 
by instantaneous 
transitions, enables oscillatory dynamics even in the absence of intrinsic limit-cycle behavior in the underlying continuous system. 
The saddle-node bifurcation (denoted SN in Fig.~\ref{fig:6}) then marks the critical value of $ I_{ext} $ beyond which 
the fixed points disappear, leaving the limit-cycle as the only 
attractor in the system. In this purely oscillatory regime, the neuron's dynamics are entirely dictated by this reset-induced periodic orbit, 
reflecting continuous spiking behavior.

\subsection{\label{sec:hardware}
Analog hardware implementation using a Memristive Spiking Neuron\protect}
In order to put this theoretical equations and solution to more concrete grounds and
to demonstrate the generality of our minimal dynamical memory model, we shall now
present the results of its hardware implementation. 
We adopt a neuromorphic analog hardware platform of spiking neurons and dynamical 
synapses  of unprecedented simplicity that was introduced recently \cite{wu2024neurosynaptic}.
The attractive feature of those hardware models is that they are realized
using solely conventional electronic components, that are affordable and widely available.
Thus, this can be of relevance for robotics and control systems based on spiking neural
networks.
Nevertheless, we should mention that the memristive spiking neuron can also
be implemented in VLSI \cite{stoliar-vlsi}, in the case that a massive number of units would need to be implemented. 

\begin{figure}
    \centering
    \includegraphics[width=1\linewidth]{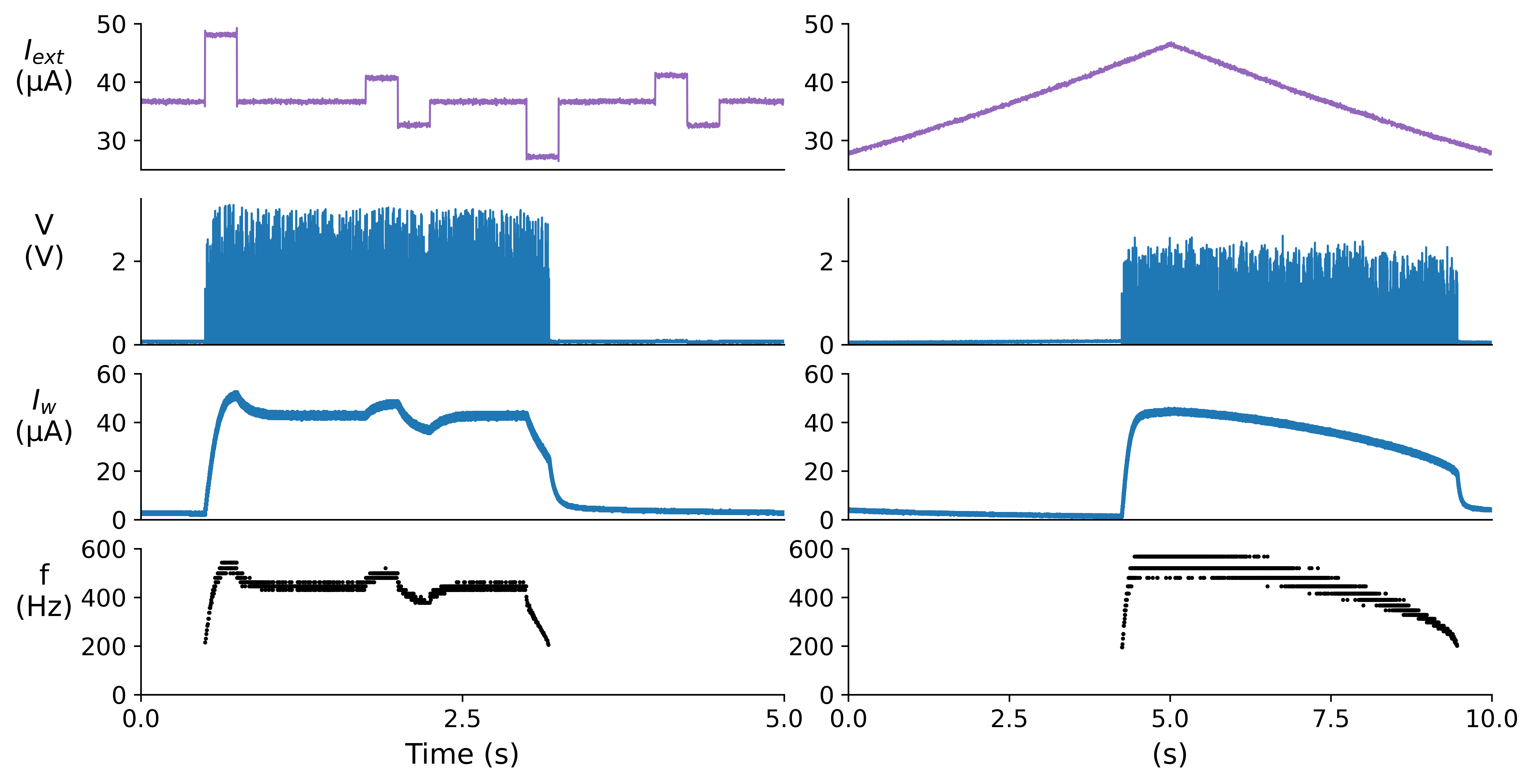}
    \caption{Left: Stability of the dynamical memory attractor in the hardware implementation. Right: Response of the same system 
    to a triangular input current. In both panels, the traces from top to bottom show: the input current, the spiking voltage, 
    the synaptic current, and the inter-spike frequency. The hardware parameters are $C$ = 33nF, $\tau_w$ = 363ms, using a P0118MA thyristor.}   
    \label{fig:hi1}
\end{figure}

Concretely, we adopt the circuit implementation for a dynamical memory model
featured in \cite{wu2024neurosynaptic}. We implemented the circuit in hardware and applied to this 
neuron with a recurrent self-excitation, similar protocols of external input current as
in Figs.~\ref{fig:2}. and \ref{fig:4}. above. The measured data are shown in Fig.~\ref{fig:hi1}.
We observe that an initial excitatory current pulse at $t$ = 0.5s and 0.25s duration is 
able to toggle the neuron from the quiescent state to the self-sustained spiking 
state with a stable frequency. At $t$ = 2s we perturb the state to demonstrate its robust stability. 
This is indeed verified as one can observe that the value of the stable frequency rapidly recovers
after the perturbation. Then, at $t$ = 3s we apply a strong inhibitory pulse, which toggles
the dynamical state back to the quiescent one. Finally, for completeness we perturb the 
inactive state to demonstrate its stability. From the comparison of the two lower panels, 
self-excitatory current $w$ and instantaneous spiking frequency $f$, we also verify
that the hardware implementation also fulfills the linearity between these
two quantities, as was discussed before for the mathematical model (Eq.~\ref{eq:lin}).

In the second set of panels, on the right, we apply protocol of linearly increasing input
current and then decreasing, to demonstrate the hysteresis effect. The measured data
are in excellent agreement with the SEQIF model results shown in Fig.~\ref{fig:4}. 

\section{\label{sec:level1}Conclusion\protect}
In this work we have introduced a minimal model of a dynamical (or working) memory. The model was 
first defined mathematically, by extending a well known model, the quadratic integrate and fire, which is
closely related to the Izhikevich model. The recurrence was introduced by just a sign change
into the model equation, turning the familiar adaptation current into a recurrent excitation.
We emphasize that the realization of this important neurocomputing primitive does not depend on
the details of the model, so it may also be implemented in other popular models such as AdEx and
Izhikevich. However, it cannot be implemented in ReLU, since its response function is exactly
linear, lacking the necessary concavity and reset mechanisms, hence it only has an unstable fixed-point. 
We analyzed the model as a self-consistent problem, which brought insight on the nature of the
stability of the fixed-points. This was verified by numerical simulations. We discussed the global
dynamical behavior by means of the nullclines construct, which brought further understanding of the
model. Finally, we implemented the model in recently introduced analog hardware and demonstrated 
that the physical embodiment of the dynamical memory is in excellent agreement with the mathematical model. 

It is worth pointing out that the present work is a significant step forward in our understanding of
spiking neural systems. In fact, the conventional wisdom, and textbooks, assume that to implement
a working memory one needs a neural population. We demonstrated here that this is not the case. Hence,
the minimal unit of dynamical memory is realized by a single neuron.

In biological systems this self-excitatory recurrence can be realize by the Ca-ionic intrinsic 
membrane currents. It is interesting to mention that the behavior of a dynamical memory is also 
analog to the persistent firing of a bump of activity, which is a feature of navigation systems.
In this regards, recent findings in the head direction system of the fruit-fly showed that a bump 
of activity is realized and evolves within the ellipsoidal body, which only counts with a handful
($\sim$ 10) of neurons \cite{noorman}. Hence, the possibility of a single neuron bump of activity is a 
relevant finding.

The extreme simplicity and frugality of the hardware implementation is an exciting feature for 
robotics and control systems. This is in stark contrast to complex VLSI CMOS ASICs and FPGAs, and 
points to the possibility of implementing cheap and reliable hardware for applications of spiking 
neuron based artificial intelligence in engineering systems.

\begin{acknowledgments}
This project has received financial support from the CNRS through the MITI interdisciplinary programs
and by from the French 
ANR ``MemAI'' project ANR-23-CE30-0040.
Centre de Calcul Intensif d’Aix-Marseille is acknowledged for granting access to its high performance computing resources. 
The preparation of this article was funded through the EU’s Horizon Europe Programme SGA 101147319 (EBRAINS 2.0).
\end{acknowledgments}

\appendix 

\section{Other non-linearities: Izhikevich and AdEx neuron models}
Similar results can be achieved using other non-linearities. In this appendix we propose two examples, 
using the exponential integrate-and-fire model and the Izhikevich model. 
\subsection{\label{app:eLIF}Exponential integrate-and-fire}

Modifying the original AdEx \cite{Brette2005-bb} , we obtained the following equations:

\begin{align*}
C \frac{dV}{dt} &= g_L (E_L - V) + g_L \Delta_T \exp\left(\frac{V - V_T}{\Delta_T}\right) + I_w + I_{\text{ext}} \\
\tau_w \frac{dI_w}{dt} &= a \, (V - E_L) - I_w
\end{align*}
With the reset condition:
\begin{align*}
    {\rm if}\ \ \ \  V = V_{spike} \ \ {\rm then}\ \ \  I_w \leftarrow I_w + b  
\end{align*}

\begin{figure}[!h]
    \centering
    \includegraphics[width=1\linewidth]{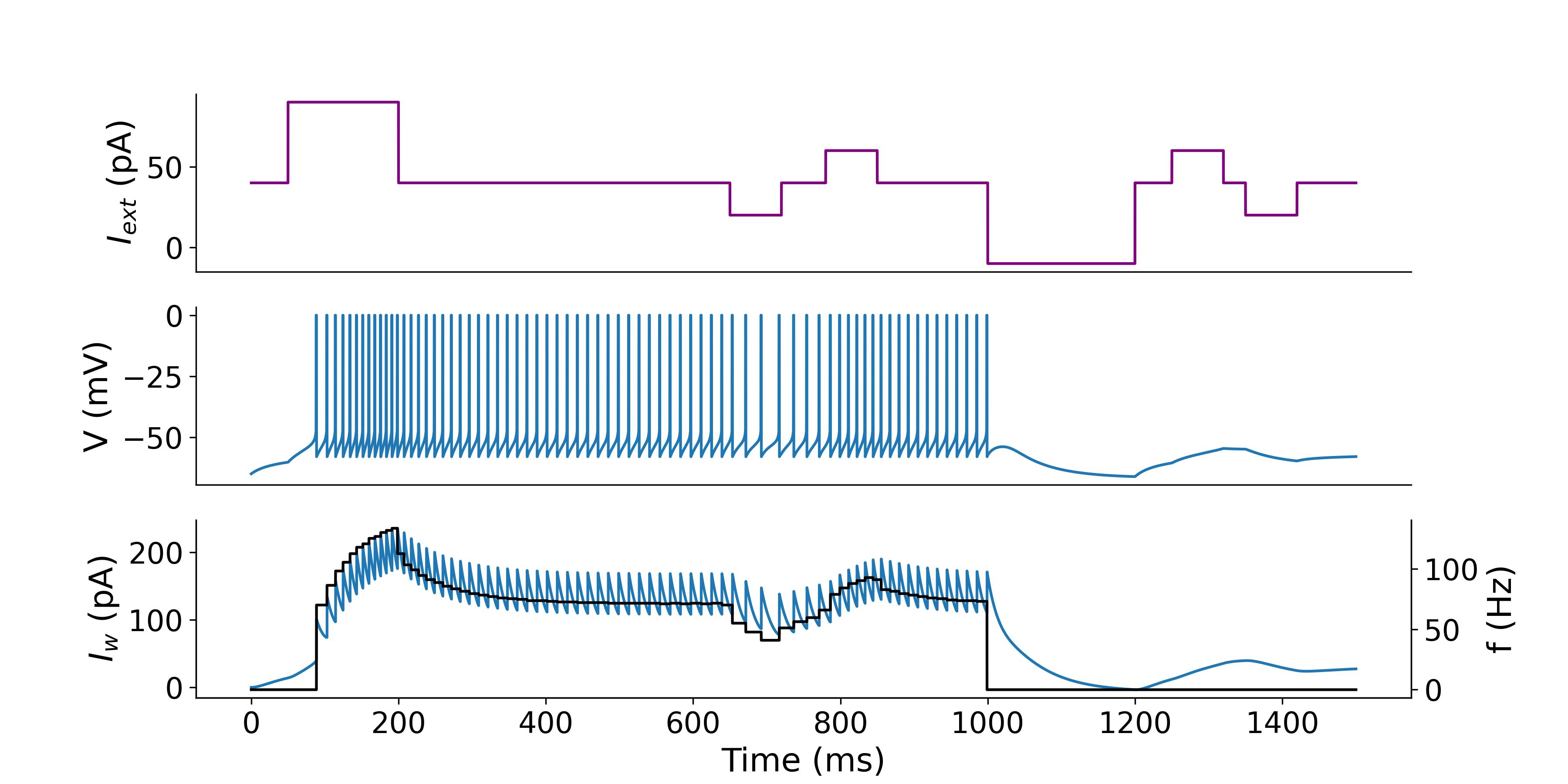}
    \caption{Response of the dynamical memory attractor based on the exponential integrate-and-fire model.}
    \label{fig:Ap1}
\end{figure}

The parameters used for Fig.~\ref{fig:Ap1} are $C = 180 \, \text{pF}$, $g_L = 10 \, \text{nS/mV}$, $E_L = -65 \, 
\text{mV}$, $V_T = -55 \, \text{mV}$, $\Delta_T = 2 \, \text{mV}$, $\tau_w = 20 \, \text{ms}$, $a = 4 \, \text{nS}$, $b = 60 \, \text{pA}$, and $V_{\text{reset}} = -58 \, \text{mV}$.

\subsection{\label{app:Imodel}Izhikevich model}

Based on the original Izhikevich model \cite{Izhikevich2003-ed}, modified with a positive $I_w$, we obtain the following equations:
\begin{align*}
\frac{dV}{dt} &= 0.04V^2 + 5V + 140 + I_w + I \\
\frac{dI_w}{dt} &= a(bV - I_w)
\end{align*}
With the reset condition:
\begin{align*}
    {\rm if}\ \ \ \  V = V_{spike} \ \ {\rm then}\ \ \  I_w \leftarrow I_w + b  
\end{align*}

\begin{figure}[!h]
    \centering
    \includegraphics[width=1\linewidth]{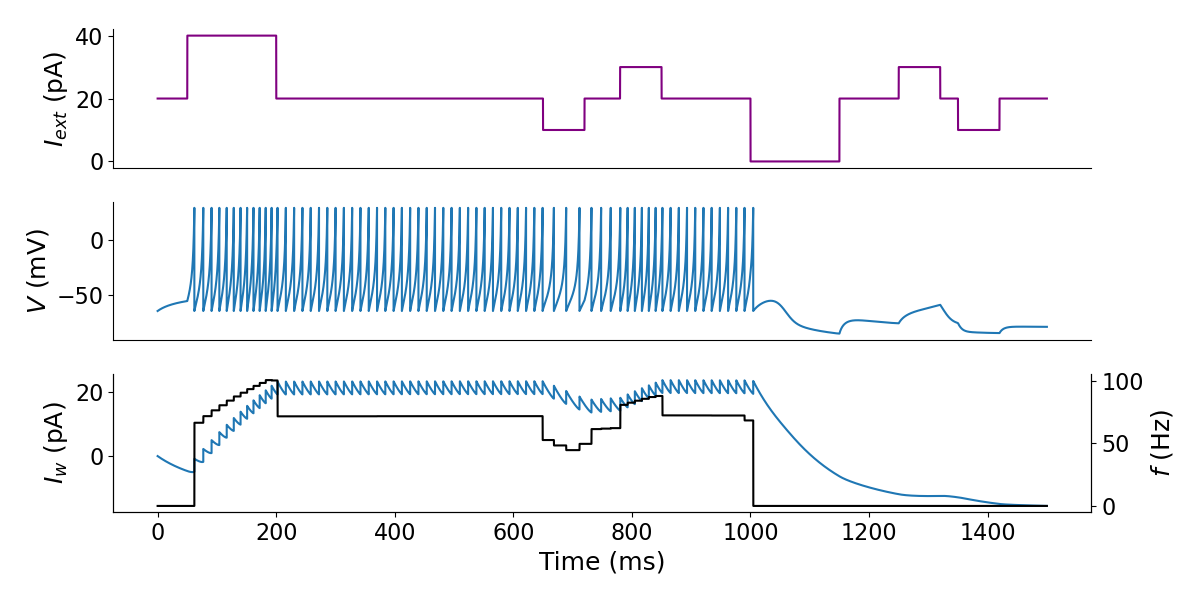}
    \caption{Response of the dynamical memory attractor based on the Izhikevich model, with $a= 0.1$ and $b=0.2$}   
    \label{fig:Ap2}
\end{figure}

We obtained qualitatively similar dynamics as shown in Fig.~\ref{fig:Ap2}.

\newpage

\bibliography{main.bib}

\begin{thebibliography}{18}%
\makeatletter
\providecommand \@ifxundefined [1]{%
 \@ifx{#1\undefined}
}%
\providecommand \@ifnum [1]{%
 \ifnum #1\expandafter \@firstoftwo
 \else \expandafter \@secondoftwo
 \fi
}%
\providecommand \@ifx [1]{%
 \ifx #1\expandafter \@firstoftwo
 \else \expandafter \@secondoftwo
 \fi
}%
\providecommand \natexlab [1]{#1}%
\providecommand \enquote  [1]{``#1''}%
\providecommand \bibnamefont  [1]{#1}%
\providecommand \bibfnamefont [1]{#1}%
\providecommand \citenamefont [1]{#1}%
\providecommand \href@noop [0]{\@secondoftwo}%
\providecommand \href [0]{\begingroup \@sanitize@url \@href}%
\providecommand \@href[1]{\@@startlink{#1}\@@href}%
\providecommand \@@href[1]{\endgroup#1\@@endlink}%
\providecommand \@sanitize@url [0]{\catcode `\\12\catcode `\$12\catcode `\&12\catcode `\#12\catcode `\^12\catcode `\_12\catcode `\%12\relax}%
\providecommand \@@startlink[1]{}%
\providecommand \@@endlink[0]{}%
\providecommand \url  [0]{\begingroup\@sanitize@url \@url }%
\providecommand \@url [1]{\endgroup\@href {#1}{\urlprefix }}%
\providecommand \urlprefix  [0]{URL }%
\providecommand \Eprint [0]{\href }%
\providecommand \doibase [0]{https://doi.org/}%
\providecommand \selectlanguage [0]{\@gobble}%
\providecommand \bibinfo  [0]{\@secondoftwo}%
\providecommand \bibfield  [0]{\@secondoftwo}%
\providecommand \translation [1]{[#1]}%
\providecommand \BibitemOpen [0]{}%
\providecommand \bibitemStop [0]{}%
\providecommand \bibitemNoStop [0]{.\EOS\space}%
\providecommand \EOS [0]{\spacefactor3000\relax}%
\providecommand \BibitemShut  [1]{\csname bibitem#1\endcsname}%
\let\auto@bib@innerbib\@empty
\bibitem [{\citenamefont {Gerstner}\ \emph {et~al.}(2014)\citenamefont {Gerstner}, \citenamefont {Kistler}, \citenamefont {Naud},\ and\ \citenamefont {Paninski}}]{gerstner}%
  \BibitemOpen
  \bibfield  {author} {\bibinfo {author} {\bibfnamefont {W.}~\bibnamefont {Gerstner}}, \bibinfo {author} {\bibfnamefont {W.}~\bibnamefont {Kistler}}, \bibinfo {author} {\bibfnamefont {R.}~\bibnamefont {Naud}},\ and\ \bibinfo {author} {\bibfnamefont {L.}~\bibnamefont {Paninski}},\ }\href@noop {} {\bibfield  {journal} {\bibinfo  {journal} {Cambridge University Press}\ } (\bibinfo {year} {2014})}\BibitemShut {NoStop}%
\bibitem [{\citenamefont {Wilson}(1999)}]{wilson}%
  \BibitemOpen
  \bibfield  {author} {\bibinfo {author} {\bibfnamefont {H.}~\bibnamefont {Wilson}},\ }\href {https://books.google.fr/books?id=yLaOkgEACAAJ} {\emph {\bibinfo {title} {Spikes, Decisions, and Actions: The Dynamical Foundations of Neuroscience}}}\ (\bibinfo  {publisher} {Oxford University Press},\ \bibinfo {year} {1999})\BibitemShut {NoStop}%
\bibitem [{\citenamefont {Seung}\ \emph {et~al.}(2000{\natexlab{a}})\citenamefont {Seung}, \citenamefont {Lee}, \citenamefont {Reis},\ and\ \citenamefont {Tank}}]{seung-autapse}%
  \BibitemOpen
  \bibfield  {author} {\bibinfo {author} {\bibfnamefont {H.~S.}\ \bibnamefont {Seung}}, \bibinfo {author} {\bibfnamefont {D.~D.}\ \bibnamefont {Lee}}, \bibinfo {author} {\bibfnamefont {B.~Y.}\ \bibnamefont {Reis}},\ and\ \bibinfo {author} {\bibfnamefont {D.~W.}\ \bibnamefont {Tank}},\ }\href@noop {} {\bibfield  {journal} {\bibinfo  {journal} {Journal of Computational Neuroscience}\ }\textbf {\bibinfo {volume} {9}},\ \bibinfo {pages} {171} (\bibinfo {year} {2000}{\natexlab{a}})}\BibitemShut {NoStop}%
\bibitem [{\citenamefont {Seung}\ \emph {et~al.}(2000{\natexlab{b}})\citenamefont {Seung}, \citenamefont {Lee}, \citenamefont {Reis},\ and\ \citenamefont {Tank}}]{seung-stability}%
  \BibitemOpen
  \bibfield  {author} {\bibinfo {author} {\bibfnamefont {H.}~\bibnamefont {Seung}}, \bibinfo {author} {\bibfnamefont {D.~D.}\ \bibnamefont {Lee}}, \bibinfo {author} {\bibfnamefont {B.~Y.}\ \bibnamefont {Reis}},\ and\ \bibinfo {author} {\bibfnamefont {D.~W.}\ \bibnamefont {Tank}},\ }\href {https://doi.org/https://doi.org/10.1016/S0896-6273(00)81155-1} {\bibfield  {journal} {\bibinfo  {journal} {Neuron}\ }\textbf {\bibinfo {volume} {26}},\ \bibinfo {pages} {259} (\bibinfo {year} {2000}{\natexlab{b}})}\BibitemShut {NoStop}%
\bibitem [{\citenamefont {Brunel}(2000)}]{brunel}%
  \BibitemOpen
  \bibfield  {author} {\bibinfo {author} {\bibfnamefont {N.}~\bibnamefont {Brunel}},\ }\href@noop {} {\bibfield  {journal} {\bibinfo  {journal} {Journal of Computational Neuroscience}\ }\textbf {\bibinfo {volume} {8}},\ \bibinfo {pages} {183} (\bibinfo {year} {2000})}\BibitemShut {NoStop}%
\bibitem [{\citenamefont {Koulakov}\ \emph {et~al.}(2002)\citenamefont {Koulakov}, \citenamefont {Raghavachari}, \citenamefont {Kepecs},\ and\ \citenamefont {Lisman}}]{koulakov2002model}%
  \BibitemOpen
  \bibfield  {author} {\bibinfo {author} {\bibfnamefont {A.~A.}\ \bibnamefont {Koulakov}}, \bibinfo {author} {\bibfnamefont {S.}~\bibnamefont {Raghavachari}}, \bibinfo {author} {\bibfnamefont {A.}~\bibnamefont {Kepecs}},\ and\ \bibinfo {author} {\bibfnamefont {J.~E.}\ \bibnamefont {Lisman}},\ }\href@noop {} {\bibfield  {journal} {\bibinfo  {journal} {Nature neuroscience}\ }\textbf {\bibinfo {volume} {5}},\ \bibinfo {pages} {775} (\bibinfo {year} {2002})}\BibitemShut {NoStop}%
\bibitem [{\citenamefont {Wu}\ \emph {et~al.}(2023)\citenamefont {Wu}, \citenamefont {Wang}, \citenamefont {Schneegans}, \citenamefont {Stoliar},\ and\ \citenamefont {Rozenberg}}]{wu2023bursting}%
  \BibitemOpen
  \bibfield  {author} {\bibinfo {author} {\bibfnamefont {J.}~\bibnamefont {Wu}}, \bibinfo {author} {\bibfnamefont {K.}~\bibnamefont {Wang}}, \bibinfo {author} {\bibfnamefont {O.}~\bibnamefont {Schneegans}}, \bibinfo {author} {\bibfnamefont {P.}~\bibnamefont {Stoliar}},\ and\ \bibinfo {author} {\bibfnamefont {M.}~\bibnamefont {Rozenberg}},\ }\href@noop {} {\bibfield  {journal} {\bibinfo  {journal} {Neuromorphic Computing and Engineering}\ }\textbf {\bibinfo {volume} {3}},\ \bibinfo {pages} {044008} (\bibinfo {year} {2023})}\BibitemShut {NoStop}%
\bibitem [{\citenamefont {Wu}\ and\ \citenamefont {Rozenberg}(2024)}]{bookchapter}%
  \BibitemOpen
  \bibfield  {author} {\bibinfo {author} {\bibfnamefont {J.}~\bibnamefont {Wu}}\ and\ \bibinfo {author} {\bibfnamefont {M.}~\bibnamefont {Rozenberg}},\ }\href@noop {} {\bibfield  {journal} {\bibinfo  {journal} {Book Chapter in Memristors-The Fourth Fundamental Circuit Element-Theory, Device, and Applications}\ } (\bibinfo {year} {2024})}\BibitemShut {NoStop}%
\bibitem [{\citenamefont {Wu}\ \emph {et~al.}(2025)\citenamefont {Wu}, \citenamefont {d'Hollande}, \citenamefont {Du},\ and\ \citenamefont {Rozenberg}}]{wu2024neurosynaptic}%
  \BibitemOpen
  \bibfield  {author} {\bibinfo {author} {\bibfnamefont {J.}~\bibnamefont {Wu}}, \bibinfo {author} {\bibfnamefont {A.}~\bibnamefont {d'Hollande}}, \bibinfo {author} {\bibfnamefont {H.}~\bibnamefont {Du}},\ and\ \bibinfo {author} {\bibfnamefont {M.}~\bibnamefont {Rozenberg}},\ }\href {https://doi.org/10.1103/PhysRevApplied.23.034030} {\bibfield  {journal} {\bibinfo  {journal} {Phys. Rev. Appl.}\ }\textbf {\bibinfo {volume} {23}},\ \bibinfo {pages} {034030} (\bibinfo {year} {2025})}\BibitemShut {NoStop}%
\bibitem [{\citenamefont {Izhikevich}(2003)}]{Izhikevich2003-ed}%
  \BibitemOpen
  \bibfield  {author} {\bibinfo {author} {\bibfnamefont {E.~M.}\ \bibnamefont {Izhikevich}},\ }\href@noop {} {\bibfield  {journal} {\bibinfo  {journal} {IEEE Trans. Neural Netw.}\ }\textbf {\bibinfo {volume} {14}},\ \bibinfo {pages} {1569} (\bibinfo {year} {2003})}\BibitemShut {NoStop}%
\bibitem [{\citenamefont {Brette}\ and\ \citenamefont {Gerstner}(2005)}]{Brette2005-bb}%
  \BibitemOpen
  \bibfield  {author} {\bibinfo {author} {\bibfnamefont {R.}~\bibnamefont {Brette}}\ and\ \bibinfo {author} {\bibfnamefont {W.}~\bibnamefont {Gerstner}},\ }\href@noop {} {\bibfield  {journal} {\bibinfo  {journal} {J. Neurophysiol.}\ }\textbf {\bibinfo {volume} {94}},\ \bibinfo {pages} {3637} (\bibinfo {year} {2005})}\BibitemShut {NoStop}%
\bibitem [{\citenamefont {G{\'o}rski}\ \emph {et~al.}(2021)\citenamefont {G{\'o}rski}, \citenamefont {Depannemaecker},\ and\ \citenamefont {Destexhe}}]{Gorski2021-aa}%
  \BibitemOpen
  \bibfield  {author} {\bibinfo {author} {\bibfnamefont {T.}~\bibnamefont {G{\'o}rski}}, \bibinfo {author} {\bibfnamefont {D.}~\bibnamefont {Depannemaecker}},\ and\ \bibinfo {author} {\bibfnamefont {A.}~\bibnamefont {Destexhe}},\ }\href@noop {} {\bibfield  {journal} {\bibinfo  {journal} {Neural Comput.}\ }\textbf {\bibinfo {volume} {33}},\ \bibinfo {pages} {41} (\bibinfo {year} {2021})}\BibitemShut {NoStop}%
\bibitem [{\citenamefont {Montbri\'o}\ \emph {et~al.}(2015)\citenamefont {Montbri\'o}, \citenamefont {Paz\'o},\ and\ \citenamefont {Roxin}}]{montbrio}%
  \BibitemOpen
  \bibfield  {author} {\bibinfo {author} {\bibfnamefont {E.}~\bibnamefont {Montbri\'o}}, \bibinfo {author} {\bibfnamefont {D.}~\bibnamefont {Paz\'o}},\ and\ \bibinfo {author} {\bibfnamefont {A.}~\bibnamefont {Roxin}},\ }\href {https://doi.org/10.1103/PhysRevX.5.021028} {\bibfield  {journal} {\bibinfo  {journal} {Phys. Rev. X}\ }\textbf {\bibinfo {volume} {5}},\ \bibinfo {pages} {021028} (\bibinfo {year} {2015})}\BibitemShut {NoStop}%
\bibitem [{\citenamefont {Chen}\ and\ \citenamefont {Campbell}(2022)}]{chen-campbell}%
  \BibitemOpen
  \bibfield  {author} {\bibinfo {author} {\bibfnamefont {L.}~\bibnamefont {Chen}}\ and\ \bibinfo {author} {\bibfnamefont {S.~A.}\ \bibnamefont {Campbell}},\ }\href@noop {} {\bibfield  {journal} {\bibinfo  {journal} {Journal of Computational Neuroscience}\ }\textbf {\bibinfo {volume} {50}},\ \bibinfo {pages} {445} (\bibinfo {year} {2022})}\BibitemShut {NoStop}%
\bibitem [{\citenamefont {Wang}\ \emph {et~al.}(2024)\citenamefont {Wang}, \citenamefont {Triebkorn}, \citenamefont {Breyton}, \citenamefont {Dollomaja}, \citenamefont {Lemarechal}, \citenamefont {Petkoski}, \citenamefont {Sorrentino}, \citenamefont {Depannemaecker}, \citenamefont {Hashemi},\ and\ \citenamefont {Jirsa}}]{virtual_brain}%
  \BibitemOpen
  \bibfield  {author} {\bibinfo {author} {\bibfnamefont {H.~E.}\ \bibnamefont {Wang}}, \bibinfo {author} {\bibfnamefont {P.}~\bibnamefont {Triebkorn}}, \bibinfo {author} {\bibfnamefont {M.}~\bibnamefont {Breyton}}, \bibinfo {author} {\bibfnamefont {B.}~\bibnamefont {Dollomaja}}, \bibinfo {author} {\bibfnamefont {J.-D.}\ \bibnamefont {Lemarechal}}, \bibinfo {author} {\bibfnamefont {S.}~\bibnamefont {Petkoski}}, \bibinfo {author} {\bibfnamefont {P.}~\bibnamefont {Sorrentino}}, \bibinfo {author} {\bibfnamefont {D.}~\bibnamefont {Depannemaecker}}, \bibinfo {author} {\bibfnamefont {M.}~\bibnamefont {Hashemi}},\ and\ \bibinfo {author} {\bibfnamefont {V.~K.}\ \bibnamefont {Jirsa}},\ }\href@noop {} {\bibfield  {journal} {\bibinfo  {journal} {National Science Review}\ }\textbf {\bibinfo {volume} {11}},\ \bibinfo {pages} {nwae079} (\bibinfo {year} {2024})}\BibitemShut {NoStop}%
\bibitem [{\citenamefont {Izhikevich}(2007)}]{izhikevich-dns-book}%
  \BibitemOpen
  \bibfield  {author} {\bibinfo {author} {\bibfnamefont {E.}~\bibnamefont {Izhikevich}},\ }\href@noop {} {\bibfield  {journal} {\bibinfo  {journal} {MIT Press}\ ,\ \bibinfo {pages} {111}} (\bibinfo {year} {2007})}\BibitemShut {NoStop}%
\bibitem [{\citenamefont {Stoliar}\ \emph {et~al.}(2022)\citenamefont {Stoliar}, \citenamefont {Akita}, \citenamefont {Schneegans}, \citenamefont {Hioki},\ and\ \citenamefont {Rozenberg}}]{stoliar-vlsi}%
  \BibitemOpen
  \bibfield  {author} {\bibinfo {author} {\bibfnamefont {P.}~\bibnamefont {Stoliar}}, \bibinfo {author} {\bibfnamefont {I.}~\bibnamefont {Akita}}, \bibinfo {author} {\bibfnamefont {O.}~\bibnamefont {Schneegans}}, \bibinfo {author} {\bibfnamefont {M.}~\bibnamefont {Hioki}},\ and\ \bibinfo {author} {\bibfnamefont {M.~J.}\ \bibnamefont {Rozenberg}},\ }\href {https://doi.org/10.1088/2399-6528/ac4e2a} {\bibfield  {journal} {\bibinfo  {journal} {Journal of Physics Communications}\ }\textbf {\bibinfo {volume} {6}},\ \bibinfo {pages} {021001} (\bibinfo {year} {2022})}\BibitemShut {NoStop}%
\bibitem [{\citenamefont {Noorman}\ \emph {et~al.}(2024)\citenamefont {Noorman}, \citenamefont {Hulse}, \citenamefont {Jayaraman}, \citenamefont {Romani},\ and\ \citenamefont {Hermundstad}}]{noorman}%
  \BibitemOpen
  \bibfield  {author} {\bibinfo {author} {\bibfnamefont {M.}~\bibnamefont {Noorman}}, \bibinfo {author} {\bibfnamefont {B.~K.}\ \bibnamefont {Hulse}}, \bibinfo {author} {\bibfnamefont {V.}~\bibnamefont {Jayaraman}}, \bibinfo {author} {\bibfnamefont {S.}~\bibnamefont {Romani}},\ and\ \bibinfo {author} {\bibfnamefont {A.~M.}\ \bibnamefont {Hermundstad}},\ }\href@noop {} {\bibfield  {journal} {\bibinfo  {journal} {Nature Neuroscience}\ }\textbf {\bibinfo {volume} {27}},\ \bibinfo {pages} {2207} (\bibinfo {year} {2024})}\BibitemShut {NoStop}%
\end{thebibliography}%

\end{document}